\documentclass[a4paper,11pt,twoside]{article}

\usepackage[latin1]{inputenc}
\usepackage[english]{babel}
\usepackage[dvips]{epsfig}
\title{\bfseries The ATLAS Tile Calorimeter Test Beam Monitoring Program}
\author{Paolo Adragna, Andrea Dotti, Chiara Roda\\
\emph{Università di Pisa e} \\ 
\emph{Istituto Nazionale di Fisica Nucleare, Sezione di Pisa}}
\date{}

\begin{document}

\maketitle

\begin{abstract}
During 2003 test beam session for ATLAS Tile Calorimeter a monitoring program has been developed to ease the setup of correct running condition and the assessment of data quality. The program has been built using the Online Software services provided by the ATLAS Online Software group. The first part of this note contains a brief overview of these services followed by the full description of Tile Calorimeter monitoring program architecture and features. Performances and future upgrades are discussed in the final part of this note.
\end{abstract}

\section{Introduction}
The monitoring program described here has been developed in the 
framework of the calibration Test Beam periods carried out at CERN
on ATLAS Tile Calorimeter. 
The ATLAS calorimetric system is a composite detector which exploits different
techniques at different rapidity regions to optimize the calorimeter 
performance while maintaining a high enough radiation resistance. 
The Tile sampling calorimeter (TileCal) is the central hadronic section of this system.
It consists of three longitudinal 
samplings of iron plates and scintillating tiles. TileCal is 
composed by one central barrel and two side (extended) barrels 
consisting of 64 modules each. During the three-year calibration program 
about $12\%$ of these modules have been exposed to test beams.

The monitoring program described here (in the following referred to as PMP) 
has been developed for the 2003 Test Beam period.
This application, based on ROOT \cite{ROOT} and on the software developed by the ATLAS
Online Software Group, allows to monitor both Tile Calorimeter modules and 
beam detector data. This program has allowed to obtain a fast setup of beam conditions 
as well as a fast check of the calorimeter data quality.

A short account of the software environment where the program has 
been developed and
a detailed description of the TileCal Monitoring Task
are given in the second and third section respectively.

\section{General feature of monitoring applications}\label{section1}

Atlas Online Software provides a number of $services$ which can be used 
to build a monitoring system~\cite{TDR}. Their main task is to carry requests 
about monitoring data (e.g. request of event blocks, request of histograms...)
from monitoring destinations to monitoring sources
and then the actual monitored data (e.g. event blocks, histograms...) back from sources 
to destinations.
Four $services$ are provided to access different types of information \cite{EMS,IS,HS,ERS}:
\begin{itemize}
\item Event Monitoring Service;
\item Information Service;
\item Histogramming Service;
\item Error Reporting System.
\end{itemize}
PMP uses the Event Monitoring Service while other services
will be included in future upgrades.

\begin{figure}[ht]
\centering
\epsfig{file=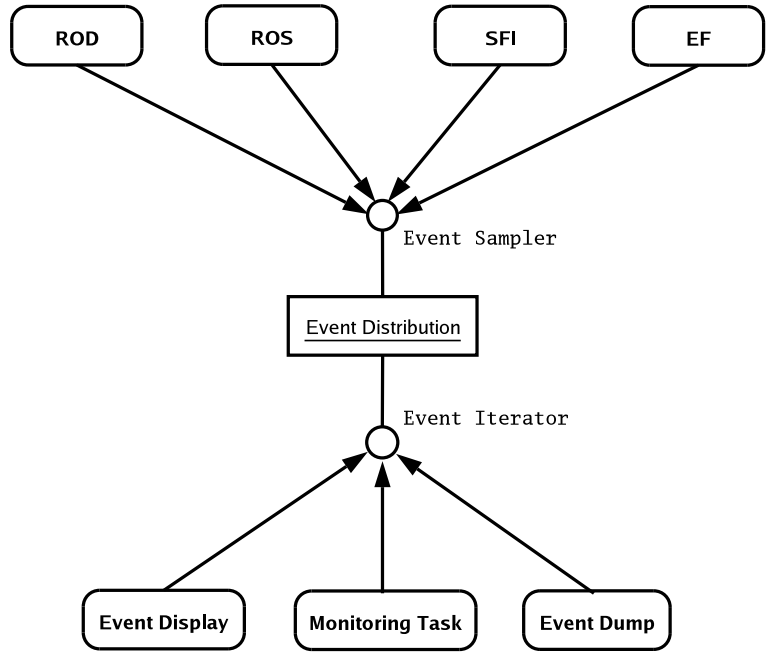,width=10cm,angle=0}
\caption{\em{\small Structure of the Event Monitoring Service~\cite{workshop}. }}
\label{fig:EMS}
\end{figure}
The Event Monitoring Service (EMS) provides events to the User Monitoring task 
sampling them from any point of the data flow chain.
The system, shown on figure \ref{fig:EMS}, consists 
of the following subsystems \cite{kolos}:

\begin{itemize}
\item the Event Sampler, which is responsible for sampling event data 
flowing through the DAQ system 
and for storing them in the Event Distribution subsystem;
\item the Event Distribution, which is responsible of the distribution, on demand;
\item the User Monitoring Task, which requests events through the Event Iterator.
\end{itemize}

The implementation of the User Monitoring Task is described in detail in the 
following sections.

\section{The TileCal monitoring task}

\begin{figure}[ht]
\centering
\epsfig{file=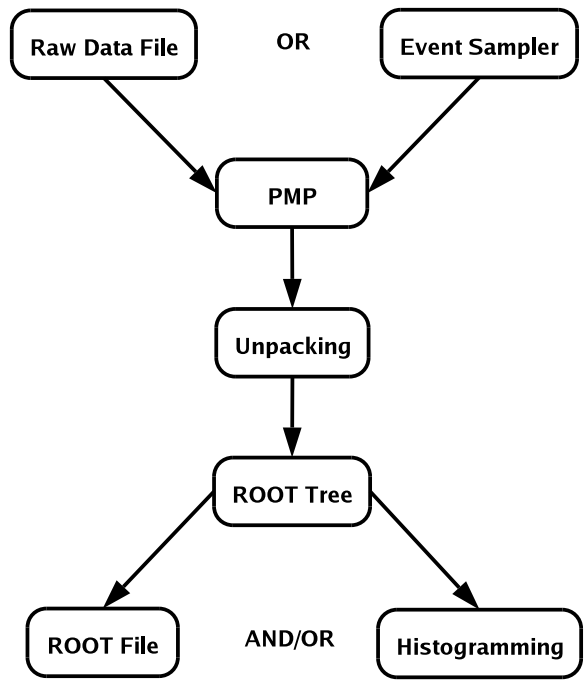,scale=0.5,angle=0}
\caption{\em{\small Schematic structure of the Tile Calorimeter monitoring program.}}
\label{fig:digforum}
\end{figure}

The TileCal Test Beam monitoring program is an object oriented application developed in C++. 
It is completely based on the ROOT framework and in particular data storage, graphical user interface and event handling fully exploit the use of ROOT classes and 
methods.
The program has been developed under Linux operating system for the i686 architecture.
In figure \ref{fig:digforum} the code working diagram is drawn \cite{digforum}.

The input to PMP may be either raw data files stored on disk
or online sampled events provided by the Event Distribution. In both cases data are
expected to be written in the standard ATLAS format~\cite{format}.

Once the event is accessed, data are unpacked and interpreted. Event unpacking
proceeds through detector independent methods up to the localization of 
the Read Out Driver (ROD) Fragments. Detector dependent methods are implemented for the extraction of data inside this fragment.
Each event contains both Tile Calorimeter and
beam detector data (Cerenkov counters, scintillators and wire chambers). 
All relevant information is extracted, event by event, and stored in a ROOT Tree \cite{guide} residing
in memory, while raw data are discarded and the buffer is freed. From this point 
on the analysis is performed using only data stored in the ROOT Tree. Histograms produced during the analysis can be immediately displayed using the presenter included inside the Graphical User Interface (GUI).

The possibility of reading raw data files from disk not only greatly simplifies the debugging process but
also allows to run simple analysis tasks on the just acquired data.

\subsection{Monitoring program architecture}

\begin{figure}[h]
\centering
\epsfig{file=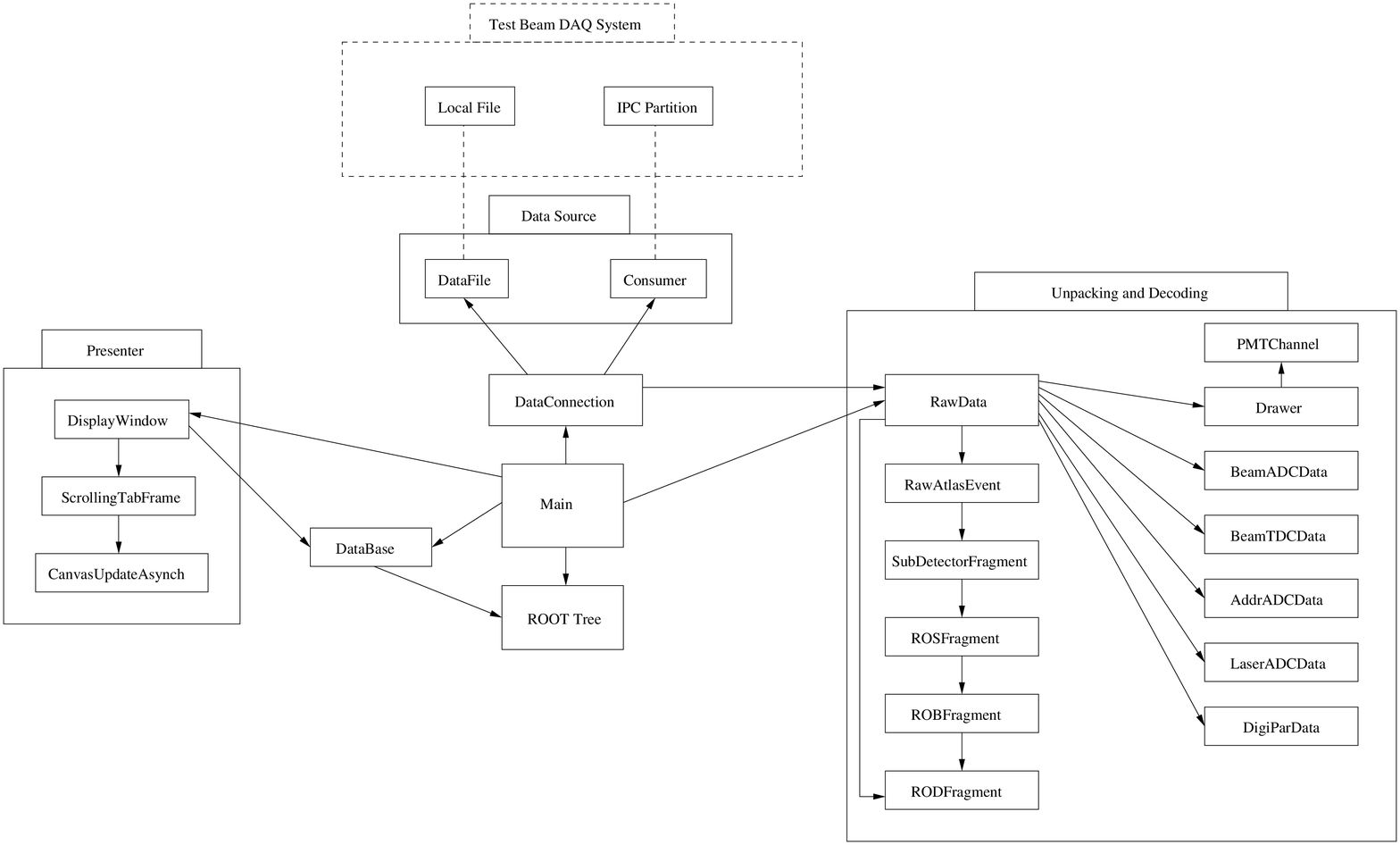,width=10cm,angle=0}
\caption{\em{\small Dependency Diagram for PMP classes.}}
\label{fig:DepDiag}
\end{figure}

The best way of explaining PMP architecture is probably to follow the path a typical user would 
walk on to work with the application \cite{help}. 
As shown in the dependency diagram of figure \ref{fig:DepDiag}, PMP abstract structure is divided into three main blocks.

The first block, named in figure as \emph{Data Source}, includes the classes \textbf{DataFile} and \textbf{Con\-sum\-er} which are responsible for fetching 
a buffer of data from the selected source.

The buffer memory address is passed to the \emph{Unpacking and Decode} block by the class \textbf{DataConnection}. The task of this second block is to extract the fragments needed and to interpret them. 

The third block, the \emph{Presenter}, fetches the histograms produced and filled by \textbf{DataBase} and organizes their graphical display into tabs. This block is also responsible for managing the user required actions.

\subsubsection*{The main panel}

\begin{figure}[ht]
\centering
\epsfig{file=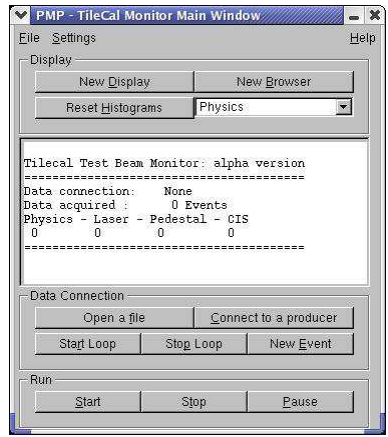,width=5cm,angle=0}
\caption{\em{\small Main control Panel of PMP.}}
\label{fig:main}
\end{figure}

When the program is started the main panel (figure \ref{fig:main})
is created.

It is built by an instance of the class \textbf{MainFrame} that, apart from drawing the GUI, is also responsible for the event 
handling and for the subsequent actions. From this stage on the user can completely drive the application 
(e.g. choose the trigger type and the data input source, opening the presenter...). 
The event access is implemented using the classes \textbf{DataFile} and \textbf{Consumer} for event retrieving either from data file or from 
online Event Monitoring Service respectively. In both cases a full event is buffered in memory and passed to the unpacking and decoding phase.

\subsubsection*{Unpacking and decoding}

The event is structured according to the ATLAS event format: the detector data is 
encapsulated into 4 layers of headers, all having similar structures. In order to 
easily reach the detector data five classes have been developed:

\begin{itemize}
\item \textbf{RawAtlasEvent}
\item \textbf{SubDetectorFragment}
\item \textbf{ROSFragment}
\item \textbf{ROBFragment}
\item \textbf{RODFragment}
\end{itemize}

Their algorithms allow to identify the corresponding block of data and their fields.
These classes are implemented in such a way that a call to the method \textbf{Raw\-Atlas\-Event::Read\-From\-Mem}
triggers the complete event decoding: in fact this method calls \textbf{Sub\-Detector\-Frag\-ment::\-Read\-From\-Mem} for every SubDetector 
block of data, every \textbf{Sub\-Detector\-Frag\-ment::\-Read\-From\-Mem} calls \textbf{ROS\-Frag\-ment::Read\-From\-Mem} for every ROS block 
and so on.

The modularity of this structure allows to easily add sub-detectors managing the decoding of their new data format 
just by including the appropriate Decoding routine.

Once the whole chain has been executed the pointers to all RODs in the event are stored in a vector.
The decoding of the detector dependent data is managed by the method \textbf{RawData::Decode} which 
determines the correct unpacking algorithm for each subsystem on the base of the ROD $id$ value 
(every ROD, and consequently every kind of data, is identified by a unique number indicated as \emph{source identifier}).

\begin{figure}[t]
\centering
\epsfig{file=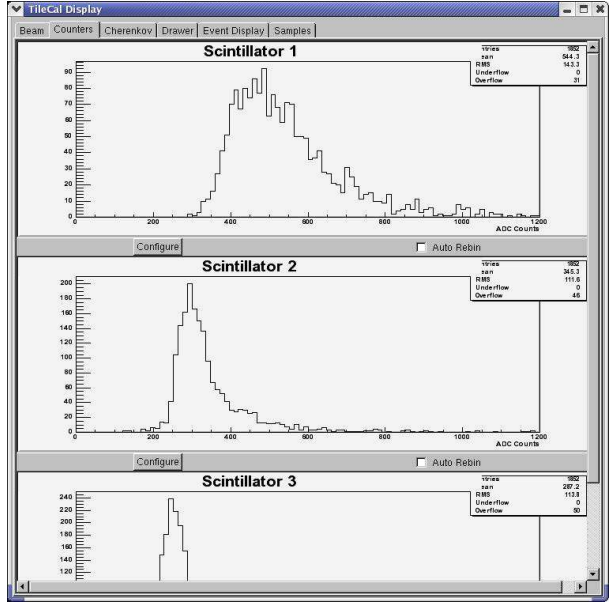,width=8cm,angle=0}
\caption{\em{\small Example of a set of PMP histograms as they appear in the Presenter.}}
\label{fig:histodisplay}
\end{figure}
  
\subsubsection*{ROOT Tree}
Once fully unpacked, data are stored in a ROOT Tree and 
the raw data buffer is freed. The tree structure is created on the base of the 
event structure. Since this one is not known before starting a 
data taking (for example the number of detectors may change in different
data takings) the tree arrangement is built according to the structure of the 
first retrieved event. The data stored in the the ROOT Tree is then used for 
simple analysis and to fill histograms.

The Tree normally acts as an interface between the data decoding and the analysis 
phase where histograms are filled; 
however it allows also to save a complete collection of sampled and preprocessed 
events on disk for a more in-depth analysis with standard ROOT tools.

\subsubsection*{Presenter}
The list of created histograms is stored inside an instance of the class \textbf{DataBase}.Histograms can be viewed with a simple presenter in a separate window. Different graphical sections
lay on the same display. Within each section, plots are drawn in different embedded canvas whose coexistence 
is granted using ROOT's capabilities of organizing graphic widgets into a \emph{table layout} \cite{guide} (figure \ref{fig:histodisplay}). The presenter refers to the class
database to get the appropriate histograms to be displayed.
The use of the \textbf{TQObject} ROOT class and the Signal -- Slot mechanism \cite{trolltech} allows to plot histograms in real time 
and to refresh each canvas independently.
\begin{figure}[t]
\centering
\epsfig{file=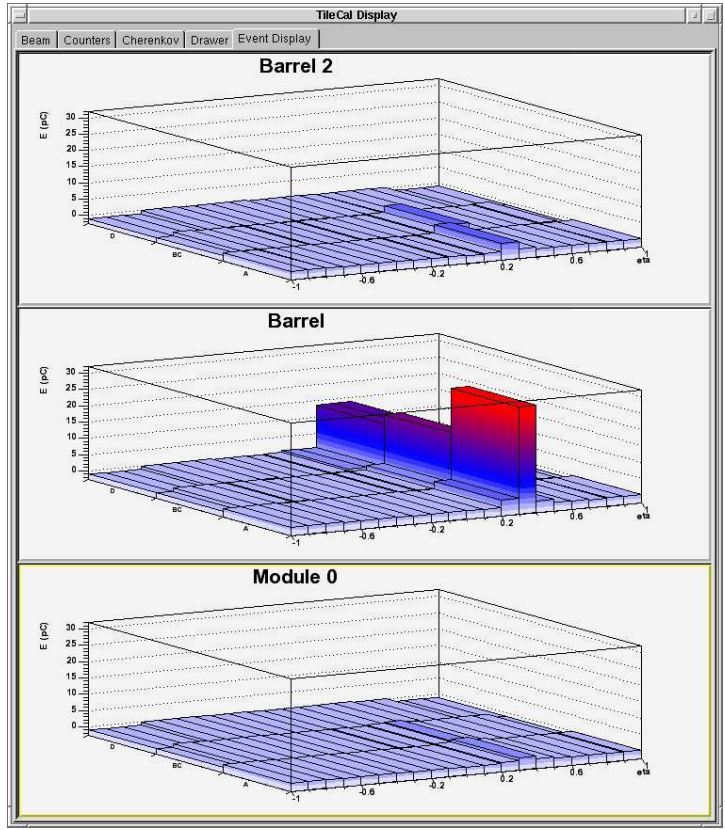,width=9.5cm,angle=0}
\caption{\em{\small Example of event display. The figure shows a representation of the energy released inside Tile Calorimeter Modules by a pion, impinging at $\eta=0.35$ and producing a signal in cells A4, BC4, D4.}}
\label{fig:eventdisplay}
\end{figure}

The Signal -- Slot mechanism has also been applied to the implementation of histogram rebinning and resetting. 
Each histogram is associated with a set of buttons, clicking each of them causes the emission of a signal. 
The latter is caught by \textbf{DataBase} which performs the appropriate action.

PMP displays plots relative to the beam detectors 
(Cerenkov counters, beam profile, coincidence scintillators) and to TileCal modules. 

Among the different plots there is a simple event display where the energy deposited in each cell of the three calorimeter modules is shown (figure \ref{fig:eventdisplay}).

Here the three modules of Test Beam setup are represented as two dimensional histograms, while each calorimeter cell is drawn as a single parallelepiped with height proportional to energy deposit. On the axes one can read the $\eta$ value and the depth (the three longitudinal layers denoted A, BC, D) of the signal.
 
On all the displayed histograms it is possible to perform standard ROOT actions (rescaling, fitting, ...).

\section{Program performances}
The monitoring program has been extensively tested to check the usage of CPU and memory during 
the data taking periods at Test Beam.
The test is performed on a PentiumIII class machine at 1GHz with 256 MB of RAM running Linux RedHat 7.3 (CERN version). 

The mean CPU load is 41\% while the average usage of physical memory is 18\%. 

\begin{figure}[ht]
\centering
\epsfig{file=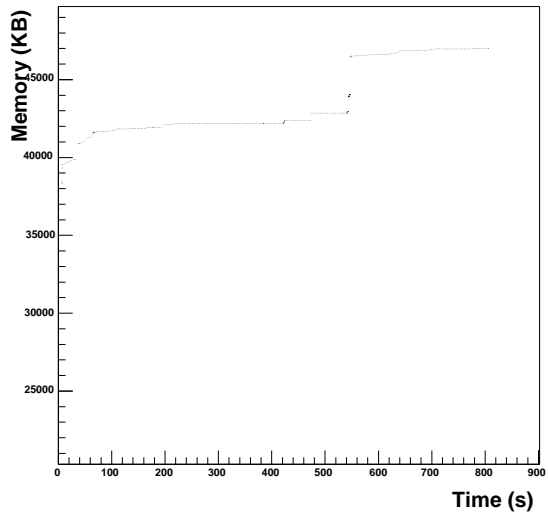,width=8cm,angle=0}
\caption{\em{\small PMP Memory usage (physical plus swap) as function of running time.}}
\label{fig:memusage}
\end{figure}

The total memory (physical plus swap) usage is shown in fig. \ref{fig:memusage} as a function of running time. 
The steep increase of used memory, shown on the plot around 550 seconds,
represents the opening of the presenter GUI. 

The monitoring of the memory usage has proved to be an important
tool to find and correct memory leaks during the development.

\section{Future plans and conclusions}
Although the program is quite flexible from the maintenance point of view, thanks to its high modularity and the use of ROOT Tree to decouple unpacking from histogramming, it still suffers from lack of optimization which limit some aspects of his performance. 

The main problem is speed. PMP keeps on making remarkable efforts to transform raw data into values ready-to-use for analysis and plotting, but his single-threaded configuration is quite penalizing. In particular situation of CPU overloading, the execution of the graphical presenter code lowers the performances considerably.

To solve this problem we plan to split the graphical part from the computational one, switching to a client-server configuration. 
The computational part which will unpack, decode and fill the required histograms, will transfer the filled histograms
to the Histogramming Service provided by the Online Software Group. The graphical part will fetch histograms from this service and will
display them. The graphical part will be configurable in order to be easily used both for different detectors and for a combined run.
\begin{figure}[ht]
\centering
\epsfig{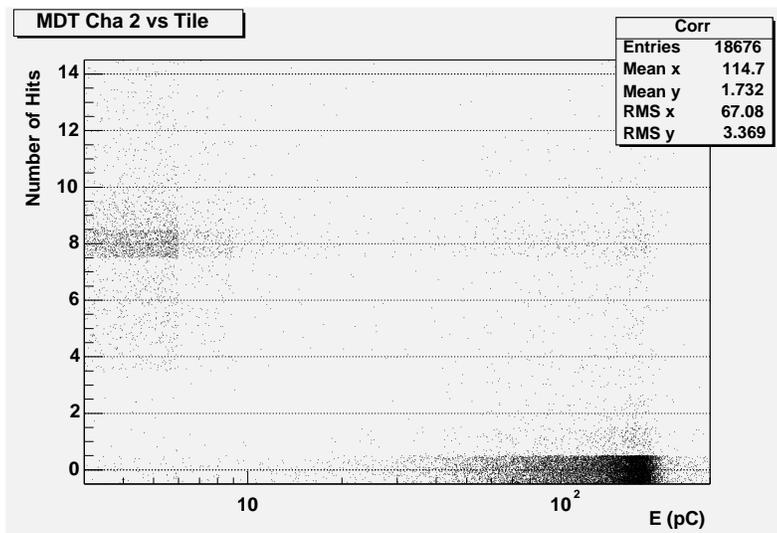}
\caption{\em{\small Number of hits recorded by the first Muon Chamber (MDT) versus 
total deposited energy in the hadronic Tile Calorimeter obtained from 
data recorded at Combined Test beam during September 2003. The two populated 
regions on the top left and on the low right regions correspond to muons
and pions respectively.}}
\label{fig:MDTvsTILECAL}
\end{figure}

A first look at the possibilities of upgrading the monitoring program here discussed to monitor a larger set of detectors 
has been tried at the combined Test Beam in September 2003. MDT and SCT were simply included using their standard data 
decoding routines. This allowed to obtain simple online histograms showing correlated information from different 
detectors (Figure~\ref{fig:MDTvsTILECAL}). 

This experience has allowed to better understand bad and good features of the present 
program and has helped us to plan the future program structures. It is foreseen to test the future versions of the monitoring 
program during the 2004 Combined Test Beam.

\section{Acknowledgments}
We would like to thank Lorenzo Santi for giving us the code he wrote for the
monitoring of the pixel detector which was the starting point of our project. We would also 
like to thank J.~E.~Garcia Navarro, S.~Gonzalez, R.~Ferrari, W.~Vandelli
for helping us with the integration of MDT and SCT detectors in the monitoring program 
during September 2003 Combined Test Beam.

\end{document}